# The Effects of Research Level and Article Type on the Differences between Citation Metrics and F1000 Recommendations

Jian Du, Xiaoli Tang, and Yishan Wu

**Jian Du:** Information Management School of Nanjing University, Nanjing, China; Institute of Medical information, Chinese Academy of Medical Sciences, Beijing, China.
**Xiaoli Tang:** Institute of Medical information, Chinese Academy of Medical Sciences, Beijing, China
**Yishan Wu:** Institute of Scientific & Technical Information of China, Beijing, China
**Corresponding author:** Xiaoli Tang

**Abstract:** F1000 recommendations have been assessed as a potential data source for research evaluation, but the reasons for differences between F1000 Article Factor (FFa scores) and citations remain unexplored. By linking recommendations for 28,254 publications in F1000 with citations in Scopus, we investigated the effect of research level (basic, clinical, mixed) and article type on the internal consistency of assessments based on citations and FFa scores. Research level has little impact on the differences between the two evaluation tools, while article type has a big effect. These two measures differ significantly for two groups: 1) *non-primary research* or *evidence-based research* are more highly cited but not highly recommended, while 2) *translational research or transformative research* are more highly recommended but gather fewer citations. This can be expected since citation activity is usually practiced by academic authors while the potential for scientific revolutions and the suitability for clinical practice of an article should be investigated from a practitioners' perspective. We conclude with a recommendation that the application of bibliometric approaches in research evaluation should consider the proportion of three types of publications: evidence-based research, transformative research, and translational research. The latter two types are more suitable for assessment through peer review.

## Introduction

Many stakeholders are concerned with how to properly assess the true impact of biomedical research. Investigators and research institutions often assess impact through such simple measures as publications in peer-reviewed journals, the impact factors of those journals, success in acquiring research grants, and the awarding of patents for novel inventions (Dembe, Lynch, Gugiu, & Jackson, 2014). The San Francisco Declaration on Research Assessment (DORA), initiated by the American Society for Cell Biology (ASCB) together with a group of editors and publishers of scholarly journals, recognizes the need to improve the methods applied to evaluate the outputs of scientific research. The general recommendation is "Do not use journal-based metrics, such as Journal Impact Factors, as a surrogate measure of the quality of individual research articles, to assess an individual scientist's contributions, or in hiring, promotion, or funding decisions" (Way & Ahmad, 2013).

Citation analysis, as one of the key methodologies in bibliometrics, has become an important tool for research performance assessment in the biomedical sciences (Du & Tang, 2013; Patel et al., 2011; Walker, Sykes, Hemmelgarn, & Quan, 2010). However, before bibliometric evaluation became widely accepted, peer review was the main tool used for research evaluation. This traditional approach received a new breath of life when the Faculty of 1000 Biology (F1000

Biology) was launched in 2002 to evaluate the quality of biomedical literature through a post-publication peer review system. F1000 Medicine was initiated in 2006, and the two services were combined in 2009 to form F1000Prime, which has built a peer-nominated global team of faculty members tasked to identify and recommend important biomedical articles (Vardell & Swogger, 2014; Wets, Weedon, & Velterop, 2003). When faculty members recommend an article, they also write a brief review explaining its importance and rate it as "Good", "Very Good" or "Exceptional" (equivalent to scores of 1, 2 or 3 stars, respectively). F1000Prime uses these individual scores to calculate a total score for each article, namely F1000 Article Factor (FFa score). F1000 recommendations have been assessed as a potential new data source for research evaluation (Waltman & Costas, 2014).

The association between citations and FFa scores has been investigated in recent years. Allen, Jones, Dolby, Lynn, and Walport (2009) compared 687 research papers assessed by Wellcome Trust reviewers with citations and F1000 ratings, and found that despite a strong positive association at an aggregate level, bibliometric measures may not be sufficient for evaluating research quality and importance, and especially when assessing individual papers or small groups of research publications. Wardle (2010) argued that within the 1,530 publications in the field of ecology, recommendations are a poor predictor of highly cited publications, which may be caused by the uneven distribution of F1000 faculty members over different areas of ecological research. Using a sample of 1,397 articles on genomics and genetics, Li and Thelwall (2012) reported a weak correlation between recommendations and citations, which measure different dimensions from different perspectives (recommendations measure the article quality from an expert perspective whereas citations measure research impact from an author perspective). Based on the data from 125 papers on cell biology or immunology from F1000 and InCites, Bornmann and Leydesdorff (2013) investigated the relationship between peer ratings and bibliometric metrics. They concluded that the correlation between recommendations and the best citation-based metrics has "at least a medium effect size." Mohammadi and Thelwall (2013) found a low but significant correlation between FFa scores and citations for a random sample of 900 clinical articles recommended by F1000. However, all the above studies were based either on relatively small samples or on just one or two specific subject categories. Waltman and Costas (2014) analyzed F1000 recommendations more comprehensively, taking into account all recommendations in the F1000 system. They found a clear correlation between F1000 recommendations and citations. However, the correlation was relatively weak, at least weaker than that between journal impact and citations. Mohammadi and Thelwall (2013) distinguished different types of publications (as indicated by labels such as "new finding", "confirmation", and "changes clinical practice" assigned by F1000 faculty members) and found a significant difference in citations and FFa scores for the two article labels: "new finding" and "changes clinical practice". They suggested that differences between recommendations and citations may relate to the type of research reported in a publication. But the interpretations of the differences require a more in-depth investigation, including analysis of the reasons F1000 faculty members give to recommend a publication, and possible biases in F1000's peer-nomination system for selecting faculty members. Waltman and Costas (2014) also suggested that more research is needed to identify the main reasons for differences between recommendations and citations in assessing the impact of publications.

Inspired by Mohammadi and Thelwall (2013) and Waltman and Costas (2014), we sought to investigate the differences between recommendations and citations from the perspectives of biomedical research level and article type (as indicated by articles labels). The concept of "research level", i.e., whether the research tends to be basic or applied/clinical, was first introduced by Narin, Pinski, and Gee (1976), who assigned 900 biomedical journals to one of four research levels, namely "Clinical Observation" (Level 1), "Clinical Mix" (Level 2), "Clinical Investigation" (Level 3), or "Basic Research" (Level 4), ranging from the most applied to the most fundamental. Articles assigned to "Clinical Mix" indicated such studies that contain a fairly even mix of observation and investigation, representing an intermediate level between medical research and clinical observation. This method has two limitations. First, it assumes all papers in a journal share a common research level. Second, the journal research level is assumed to remain constant, even if in reality it changes over time. To address these deficiencies, Lewison and Paraje (2004) created their own method for classifying the research level of biomedical research journals as clinical, basic, or somewhere in between. Their classification is based on counting articles that contain one of about 100 "clinical" title words, or one of a similar number of "basic" title words, or both. This allows journals to be classified rapidly and transparently, and also allows for changes in their research level over time, as well as for classification of individual papers in mixed journals as clinical or basic. Translational research, which is aimed at reducing the time required for basic discoveries to be translated into effective patient treatment, has become a topic of great interest in the biomedical community over the past several years. Much of the focus is on the linkages between basic biomedical research and clinical medicine. Subsequently, research level has been used to characterize translational pathways in biomedical research (Cambrosio, Keating, Mercier, Lewison, & Mogoutov, 2006; Jones, Cambrosio, & Mogoutov, 2011) and to assess the translational capacity of medical researchers and teams based on the research level of a researcher's output (Boyack, Klavans, Patek, Yoon, & Ungar, 2013). F1000 uses its own field classification system, which differs from the journal-based subject categories used by Web of Science or Scopus. In this study, we classify research articles into three levels: basic, clinical and mixed, according to the specialty of F1000 recommender. If an article is recommended only by faculty members specialized in biology-related subjects (scientists), it is classified as "Basic Research". If an article is recommended only by faculty members in medicine-related subjects (clinicians), it is classified as "Clinical Research". Finally, if an article is recommended by faculty members specialized in an area that involves both biological and medical research topics, or if it is rated by both scientists and clinicians, then it is classified as "Mixed Research"; this last classification thus denotes articles of interest to both basic and clinical researchers. Previous classifications of biomedical research levels were based either on predetermined sets of biological versus clinical journals (Lewison & Paraje, 2004; Opthof, 2011; Seglen, 1997), or on keywords/MeSH terms related to basic research or clinical research (van Eck, Waltman, van Raan, Klautz, & Peul, 2013; Fajardo-Ortiz, Duran, Moreno, Ochoa, & Castano, 2014a, 2014b). This study instead assigns publications to basic, clinical or mixed research according to the research fields of faculty members who recommend them. The concept of research level as operationalized in this study is validated in the following section.

Faculty members also tag articles with one or more of the following labels: 1) *Changes Clinical Practice*: the article recommends a complete, specific and immediate change in practice by clinicians for a defined group of patients; 2) *Confirmation*: the article validates previously published data or hypotheses; 3) *Controversial*: the article challenges established dogma; 4) *Good for Teaching*: a key article in the field and/or one which is well written; 5) *Interesting Hypothesis*:

the article presents a new model; 6) *New Finding*: the article presents original data, models or hypotheses; 7) *Novel Drug Target*: the article suggests new targets for drug discovery; 8) *Refutation*: the article disproves previously published data or hypotheses; and 9) *Technical Advance*: the article introduces a new practical/theoretical technique, or a novel use of an existing technique. F1000 further assigns other document types: *Clinical Trial*, *Systematic Review/Meta-analysis* and *Review/Commentary*. We primarily seek to answer two fundamental research questions:

(1) Does research level (basic, clinical or mixed) impact the association between citations and FFa scores?

(2) Do different article types impact the association between citations and FFa scores?

## Materials and methods

The subject category in F1000 is a three level system that includes subjects, sections and topics. The reviewer team comprises an international advisory board, faculty heads, section heads, and faculty members, together with their associate faculty members. The international advisory board assists with the selection of the faculty heads for each subject. Faculty heads divide their subjects into major sections and help nominate section heads. Section heads are acknowledged experts in their fields and are responsible for dividing their sections into topics and for selecting suitable faculty members to review the literature on each topic area. The number of faculty members is broadly linked to the total number of articles published in the various topic areas.

*Sample Selection*

Articles recommended by faculty members are assigned to his/her discipline, organized into 24 biology-related and 20 medicine-related subjects. As is shown in Table 1, each medicine-related subject has its own separate faculty heads, section heads or faculty members, yet this independent reviewer system does not exist in nearly half of all biology-related subjects (11/24). The number of recommended articles was unevenly distributed across biology-related subjects relative to medicine-related subjects. First, more than two-thirds of articles (80,094/115,565, 67.7%) are recommended by faculty members specialized in biology-related subjects. This proportion is much larger than that for medicine-related subjects (45,515/115,565, 38.1%). Second, within biology-related subjects, the top three subject categories in which articles were frequently recommended were cell biology, molecular medicine, and genomics and genetics, accounting for 41.8%, 31.2% and 27.4%, respectively, higher than the percentages for cancer biology, plant biology, cardiovascular biology, gastrointestinal biology, and so on. In comparison, articles are relatively balanced in their coverage of most medicine-related subjects, with each subject typically accounting for around 3–6% of all articles.

The sum of the papers under all subjects (468,928) was almost four times the total number of biology and medicine articles (115,565). A single article thus may belong to multiple subjects if it is recommended by faculty members from different subjects, so the papers assigned to a subject do not necessarily only reflect the activities of faculty members specialized in that category. Among the 44 total subjects, several pairs represent biological and medical research on the same disease area, such as Cancer Biology and Oncology, Cardiovascular Biology and Cardiovascular Disorders, Neuroscience and Neurological Disorders, Respiratory Biology and Respiratory Disorders, or Renal Biology and Nephrology. Given the complicated recommendation activities of reviewers and the uneven distribution of articles across different subjects, as well as the objectives of this study being to investigate research level in the continuum from basic to clinical research, three biology-medicine subject pairs with a relative large number of articles were selected for study: Cancer Biology, Oncology, Cardiovascular Biology, Cardiovascular Disorders, Neuroscience and Neurological Disorders.

Table 1. Distribution of reviewers and articles (published from 1999 to 2013) across biology- and medicine-related subjects

| F1000 Biology | reviewers | articles | | F1000 Medicine | reviewers | articles | |
|---|---|---|---|---|---|---|---|
| | | *N* | % | | | *N* | % |
| Cell Biology | 1437 | 49448 | 42.8 | Neurological Disorders | 863 | 7442 | 6.4 |
| Molecular Medicine | N/A | 36022 | 31.2 | Oncology | 411 | 6731 | 5.8 |
| Genomics & Genetics | 574 | 31640 | 27.4 | Public Health & Epidemiology | 282 | 6300 | 5.5 |
| Biotechnology | N/A | 27395 | 23.7 | Infectious Diseases | 529 | 6107 | 5.3 |
| Physiology | 612 | 24204 | 20.9 | Anesthesiology & Pain Management | 886 | 5851 | 5.1 |
| Biochemistry | 3 | 23438 | 20.3 | Gastroenterology & Hepatology | 632 | 5009 | 4.3 |
| Pharmacology & Drug Discovery | 809 | 21792 | 18.9 | Cardiovascular Disorders | 432 | 4832 | 4.2 |
| Structural Biology | 708 | 19918 | 17.2 | Diabetes & Endocrinology | 385 | 4643 | 4.0 |
| Microbiology | 866 | 16853 | 14.6 | Critical Care & Emergency Medicine | 328 | 4495 | 3.9 |
| Developmental Biology | 624 | 16622 | 14.4 | Respiratory Disorders | 492 | 4462 | 3.9 |
| Neuroscience | 1195 | 16542 | 14.3 | Psychiatry | 512 | 4236 | 3.7 |
| Chemical Biology | 497 | 16340 | 14.1 | Rheumatology & Clinical Immunology | 581 | 3955 | 3.4 |
| Bioinformatics & Computational Biology | N/A | 15331 | 13.3 | Women's Health | 291 | 3944 | 3.4 |
| Immunology | 986 | 13157 | 11.4 | Hematology | 530 | 3303 | 2.9 |
| Molecular Biology | N/A | 12907 | 11.2 | Dermatology | 495 | 3181 | 2.8 |
| Evolutionary Biology | N/A | 10500 | 9.1 | Urology | 325 | 3018 | 2.6 |
| Cancer Biology | N/A | 9169 | 7.9 | Nephrology | 315 | 2432 | 2.1 |
| Plant Biology | 447 | 6832 | 5.9 | Otolaryngology | 266 | 1868 | 1.6 |
| Ecology | 620 | 6583 | 5.7 | Ophthalmology | 332 | 890 | 0.8 |
| Metabolic & Endocrine Science | N/A | 3006 | 2.6 | Research Methodology | 96 | 796 | 0.7 |
| Cardiovascular Biology | N/A | 3001 | 2.6 | | | | |
| Gastrointestinal Biology | N/A | 2080 | 1.8 | | | | |
| Renal Biology | N/A | 1344 | 1.2 | | | | |
| Respiratory Biology | N/A | 1309 | 1.1 | | | | |

N/A: This subject was created by grouping relevant sections from other subjects, and thus, does not have separate Faculty Heads, Section Heads or Faculty Members.

### *Data Collection*

In this study, F1000's advanced search was used to select articles published between 1999 and 2013 and recommended by faculty members in the three subject pairs mentioned above. We captured the bibliographic and F1000 information for each article, including author(s), title, journal, publication date, FFa score, and article labels. Thomson Reuters Journal Citation Report (JCR) for 2013 was used to collect journal impact factors. Scopus was used to search citations for each article because of the following: 1) Previous studies have found that oncology or general medicine articles published from 1996 onward received more citations from Scopus than from the Web of Science (Lewison & Paraje, 2004; Seglen, 1997), and Scopus covered clinical research journals better than Web of Science does (Opthof, 2011). 2) F1000 states that only about 15% of its articles come from high-profile popular journals such as *Science* and the *New England Journal of Medicine*, with the remaining 85% coming from a wider range of more specialized and less popular journals (Vardell & Swogger, 2014). We thus use Scopus for citation matching to capture the wider citation impact.

*Dataset 1* To evaluate the general distribution of articles across different subjects and journals, the analysis included all articles published between 1999 and 2013, and recommended by F1000 faculty members specialized in the abovementioned six subjects, a total of 47,446 articles.

*Dataset 2* A link between listings in the F1000 dataset and the Scopus database was established if the publications shared either the same digital object identifier (DOI) or the same journal title, volume number, issue number, and first author name (i.e., last name and initials). Some scholars have claimed that reliably determining the citation impact of papers requires a sufficiently wide citation window of at least 3 years (Cambrosio et al., 2006), and generally more than 80% of first-citations are accrued within an initial 3-year window and more than 90% within an initial 5-year citation window (Glänzel, Schlemmer, & Thijs, 2003). Thus, a 3-5 year citation window for the most recently published articles was adopted. We only selected the 31461 articles published between 1999 and 2010 for citation matching. The citation data was retrieved on August 2014. F1000 conference posters ($n$=44) and the retracted articles ($n$=43) were excluded, leaving 31374 articles. To establish a link between individual papers and bibliometric data, two procedures were used. 1) A total of 29,361 papers in the dataset could be matched with a paper in the Scopus database using DOI. 2) For 1279 of the 2013 remaining papers, no match was achieved with DOI, but a match could be achieved for the title, journal, volume and issue. Citations in Scopus were then drawn for 30640 of the total 31374 papers (97.6%). This percentage exceeds 91% obtained by Bornmann (2014) and the 93% quoted by Waltman and Costas (2014), who used a similar procedure to match data from F1000 with bibliometric data in Web of Science. Because one article may be classified under multiple subjects, it may be recommended by faculty members specialized in different subjects. Therefore we further examined the primary data collection procedure and deleted 2386 duplicated articles recommended by faculty members affiliated with two or more of the six subjects, leaving 28,254 articles in the sample. Dataset 2 is a subset of dataset 1.

## Results

The analytical results are presented in three subsections. We first provide general statistics on: 1) the distribution of articles across subjects and journals; 2) the association between FFa scores and the other three citation measures; and 3) the content of articles across research levels as reflected by article labels, journals in which they were published and terms associated with them. We then investigate the effect of research level and article type on the internal consistency of assessments based on citations and FFa scores.

### *General Statistics*

*Distribution of articles across subjects and journals* Figure 1 shows the annual number of recommended papers in these three subject pairs. Because F1000 Biology was launched in 2002 and F1000 Medicine in 2006, the annual number of recommended papers on biology-related subjects rises early and then remains relatively stable, while that for medicine-related subjects displays a significant increase.

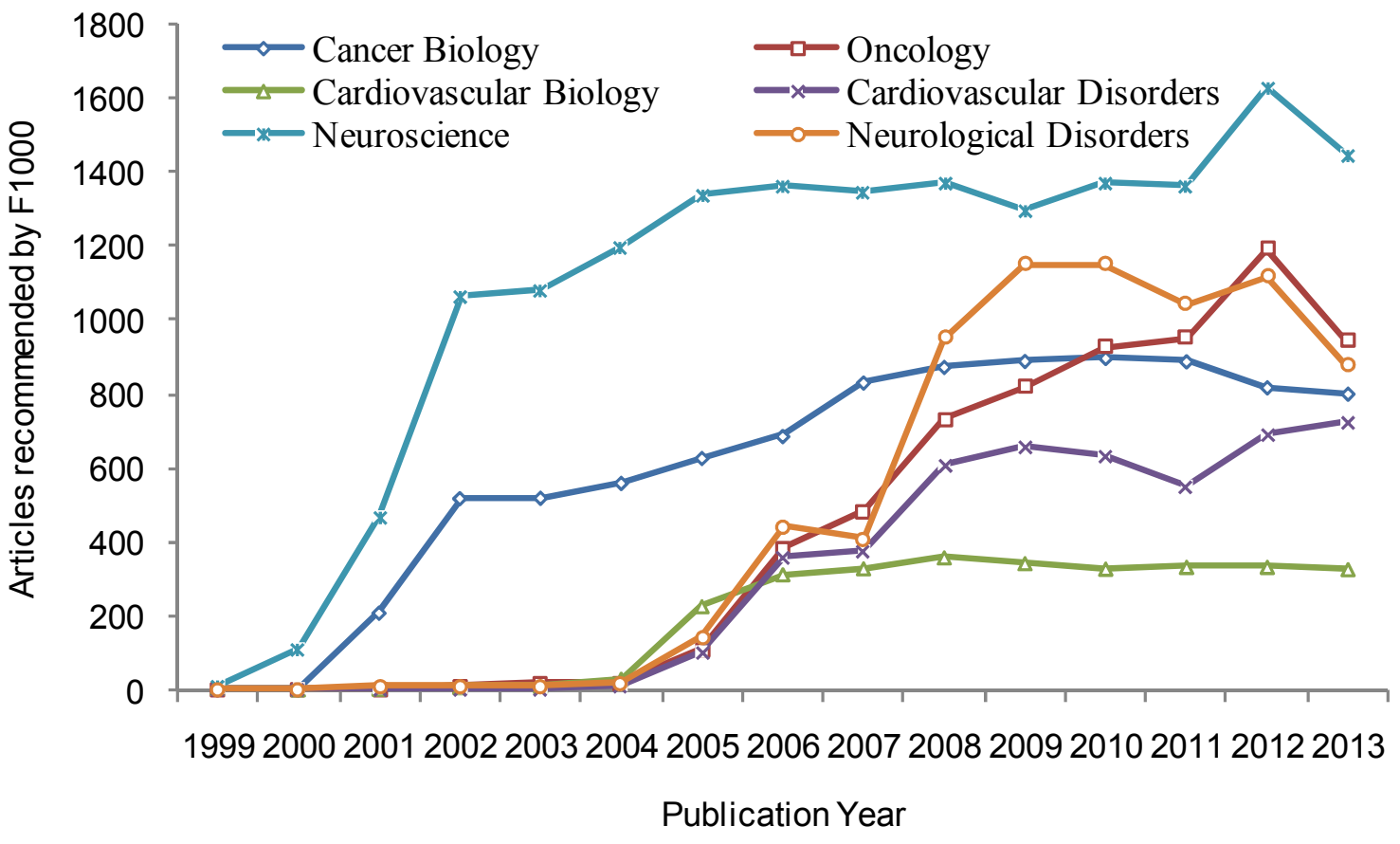

Figure 1. Annual number of recommended papers, 1999-2013.

Of the 1,904 journals in Dataset 1, 311 were not indexed by Web of Science. In Table 2, nearly 30% of cancer biology-related articles came from very high profile journals such as *Nature*, *PNAS*, *Cell* and *Science*, while the largest sources for the other five subjects were more specialized journals (e.g., *Journal of Clinical Oncology* and *Blood* for Oncology, *Circulation Research* and *Circulation* for Cardiovascular Biology and Cardiovascular Disorders, and *Journal of Neuroscience* and *Neurology* for Neuroscience and Neurological Disorders).

Table 2. Top 10 journals in each discipline by number of articles

| Journal | No. | % | Journal | No. | % |
|---|---|---|---|---|---|
| **Cancer Biology** | | | **Oncology** | | |
| *Nature* | *772* | *8.4* | *J Clin Oncol* | *652* | *9.8* |
| *Proc Natl Acad Sci U S A* | *642* | *7.0* | *Blood* | *307* | *4.6* |
| *Cell* | *630* | *6.9* | *N Engl J Med* | *233* | *3.5* |
| *Science* | *529* | *5.8* | *Nature* | *142* | *2.1* |
| *J Cell Biol* | *429* | *4.7* | *Cancer Res* | *142* | *2.1* |
| *Mol Cell* | *370* | *4.0* | *Lancet Oncol* | *134* | *2.0* |
| *Nat Cell Biol* | *342* | *3.7* | *J Urol* | *122* | *1.8* |
| *J Biol Chem* | *321* | *3.5* | *Ann Oncol* | *119* | *1.8* |
| *Genes Dev* | *308* | *3.4* | *Proc Natl Acad Sci U S A* | *115* | *1.7* |
| *Cancer Cell* | *306* | *3.3* | *Cancer* | *108* | *1.6* |
| **Cardiovascular Biology** | | | **Cardiovascular Disorders** | | |
| *Circ Res* | *187* | *6.2* | *Circulation* | *502* | *10.5* |
| *Proc Natl Acad Sci U S A* | *178* | *5.9* | *J Am Coll Cardiol* | *381* | *8.0* |
| *Circulation* | *149* | *5.0* | *N Engl J Med* | *304* | *6.4* |
| *Hypertension* | *132* | *4.4* | *JAMA* | *136* | *2.8* |
| *Nature* | *131* | *4.4* | *Lancet* | *128* | *2.7* |
| *J Clin Invest* | *109* | *3.6* | *Eur Heart J* | *112* | *2.3* |
| *Blood* | *89* | *3.0* | *Am J Cardiol* | *105* | *2.2* |
| *J Biol Chem* | *79* | *2.6* | *J Vasc Surg* | *98* | *2.0* |
| *Am J Physiol Heart Circ Physiol* | *70* | *2.3* | *Hypertension* | *90* | *1.9* |
| *Nat Med* | *69* | *2.3* | *J Am Soc Nephrol* | *81* | *1.7* |
| **Neuroscience** | | | **Neurological Disorders** | | |
| *J Neurosci* | *1894* | *11.5* | *Neurology* | *462* | *6.3* |
| *Neuron* | *1341* | *8.1* | *J Neurosci* | *258* | *3.5* |
| *Proc Natl Acad Sci U S A* | *1184* | *7.2* | *Brain* | *248* | *3.4* |
| *Nature* | *1116* | *6.8* | *Stroke* | *210* | *2.8* |
| *Nat Neurosci* | *1031* | *6.3* | *Proc Natl Acad Sci U S A* | *188* | *2.5* |
| *Science* | *917* | *5.6* | *Nature* | *159* | *2.2* |
| *Cell* | *544* | *3.3* | *Ann Neurol* | *158* | *2.1* |
| *Development* | *387* | *2.3* | *N Engl J Med* | *157* | *2.1* |
| *J Biol Chem* | *314* | *1.9* | *Pain* | *149* | *2.0* |
| *Curr Biol* | *298* | *1.8* | *Science* | *133* | *1.8* |

*Association between FFa scores and the other three citation measures* Within dataset 2, we first analyzed the distribution of FFa scores and citations for 28,254 articles, and found that the statistical curves for both FFa scores and citations follow a long-tailed distribution. In other words, there are many papers with few citations and low FFa scores. The majority of the studied articles had FFa scores of only 1 or 2, with these two categories comprising 51% (n=14,421) and 28% (n=7,852) of the total, respectively. Since the citations and FFa scores of these 28,254 articles do not follow a normal distribution, skewed distributions should not be studied in terms of central tendency statistics such as arithmetic means, but rather using nonparametric statistics, such as the top 1%, top 10%, etc. Hence we examined the distribution pattern of citations and FFa scores in terms of five percentiles, namely top 0.1%, top 1%, top 10%, top 20%, and top 50%. Table 3 displays data on the minimum number of citations and FFa scores needed to meet these five percentile breakdowns in the dataset.

Table 3. Minimum number of citations and FFa scores needed for inclusion in five percentile categories.

| Percentiles | *N* | FFa Scores | *N* | Citations |
|---|---|---|---|---|
| Top 0.1% | 29 | 20 | 29 | 2597 |
| Top 1% | 256 | 11 | 283 | 962 |
| Top 10% | 2156 | 5 | 2822 | 265 |
| Top 20% | 5941 | 3 | 5663 | 160 |
| Top 50% | 13833 | 2 | 14196 | 62 |

The Journal Impact Factor is the average number of citations accrued in the JCR year by articles from the given journal published in the past two years. The Immediacy Index is the average number of times an article is cited in the year it is published, indicating how quickly articles in a journal are cited. In general, there is only a medium level of association between FFa scores and the other three citation measures, namely Journal Impact Factor (Spearman $r = 0.428$, $p < 0.01$), Journal Immediacy Index (Spearman $r = 0.410$, $p < 0.01$) and total citations (Spearman $r = 0.363$, $p < 0.01$). In the next section, we will investigate the causes of the differences between FFa scores and citations.

*Content of articles across research levels* In our sample, the numbers of papers classified as basic, clinical or mixed research from the three biology-medicine subject pairs are 18,493, 9,069 and 692, respectively. We validate the concept of research level as operationalized from the perspective of recommender' research areas and subjective concerns, by examining the objective and natural content features of articles across research levels through article labels, journals in which they appeared, and terms.

*a) Article labels.* Since each paper will be given at least one label, the distribution of article labels across research levels was first examined (Figure 2). More than half of the articles assigned to "Novel Drug Target", "Technical Advance", "New Finding", "Interesting Hypothesis" and "Controversial" were basic, and 80% or more of the articles labeled "Changes Clinical Practice" (96%), "Randomized Controlled Trial (RCT) Clinical Trial" (85%), "Systematic Review/Meta-analysis" (84%) and "Non-RCT Clinical Trial" (84%) were clinical, clearly separating the basic with clinical research articles. In addition, all "General Interest", 15% of "Good for Teaching," and 11% of "Novel Drug Target" publications were mixed research articles. These articles, to some extent, attracted attention from both scientists and clinicians and may provide insights into the translation of research to practice.

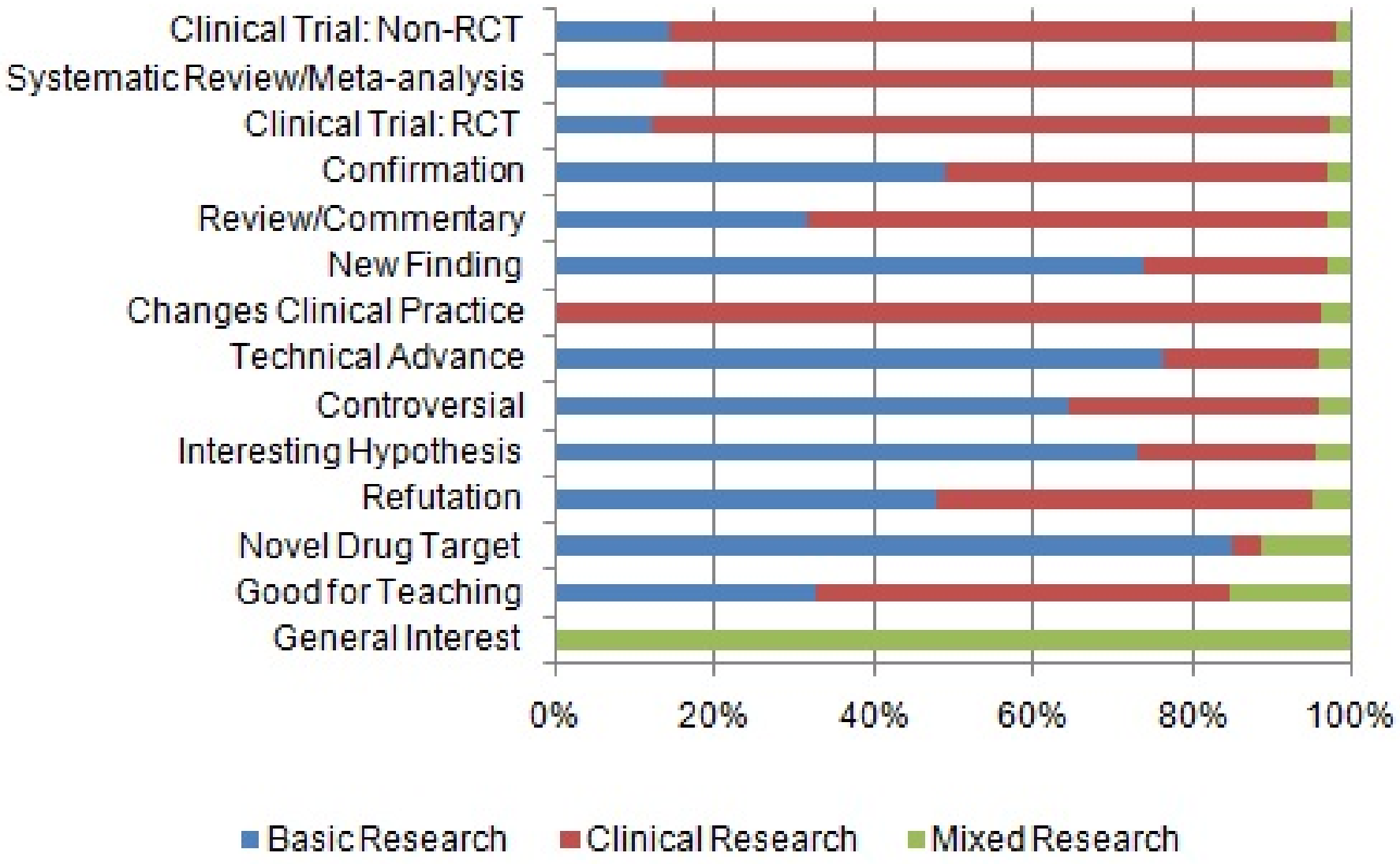

Figure 2. Distribution of article labels across research levels

*b) Journals.* The distribution of journals across research levels was then analyzed. As is shown in Table 4, more than half of basic research articles were published in either basic science journals (e.g., *Journal of Neuroscience*, *Neuron*, *Cell*, *Journal of Cell Biology*) or multidisciplinary journals (e.g., *PNAS*, *Nature* or *Science*). The top ten journals publishing the most clinical articles were all clinical-research specific, including *Journal of Clinical Oncology*, *New England Journal of Medicine*, *Circulation*, *Neurology*, *Journal of the American College of Cardiology* and *Blood*. More than two-thirds of basic-clinical mixed research articles were published in multidisciplinary journals (*Nature*, *Science* and *PNAS*), high-profile clinical or basic science journals (*New England Journal of Medicine* and *Cell*), or relevant translational research journals (*Nature Medicine*, *Nature Neuroscience*, *Nature Genetics*, *Cancer Cell* and *Journal of Clinical Investigation*). We investigated the distribution of terms extracted from 692 mixed research articles in detail.

Table 4. Distribution of journals across research levels

| Basic Research | | Clinical Research | | Mixed Research | |
|---|---|---|---|---|---|
| Journal | % of 18493 | Journal | % of 9069 | Journal | % of 692 |
| *J Neurosci* | 7.6 | *J Clin Oncol* | 4.0 | *Nature* | 14.0 |
| *PNAS* | 7.4 | *N Engl J Med* | 3.7 | *Science* | 12.1 |
| *Nature* | 6.9 | *Circulation* | 3.3 | *N Engl J Med* | 5.9 |
| *Neuron* | 5.6 | *Neurology* | 3.2 | *Nat Med* | 5.5 |
| *Science* | 5.5 | *J Am Coll Cardiol* | 2.3 | *PNAS* | 5.3 |
| *Cell* | 4.4 | *Blood* | 2.2 | *Nat Neurosci* | 5.2 |
| *Nat Neurosci* | 4.1 | *PNAS* | 1.9 | *Cell* | 4.8 |
| *J Cell Biol* | 2.8 | *Brain* | 1.8 | *J Neurosci* | 3.6 |
| *J Biol Chem* | 2.8 | *Lancet* | 1.8 | *Nat Genet* | 3.5 |
| *Curr Biol* | 2.3 | *JAMA* | 1.7 | *Cancer Cell* | 2.6 |
| *Development* | 2.0 | | | *Neuron* | 2.6 |
| | | | | *J Clin Invest* | 2.0 |
| | | | | *Circulation* | 1.6 |
| | | | | *Lancet* | 1.4 |
| Accumulated percentage | 50 | | 25 | | 70 |

*c) Terms* To analyze the content of 692 articles we extracted terms (defined as a sequence of nouns and adjectives ending with a noun) from their titles and abstracts using VOSviewer (http://www.vosviewer.com/). Two terms are said to co-occur in a publication if they both occur at least once in the title or abstract. The larger the number of publications in which two terms co-occur, the stronger the relationship between those terms. We chose the top 203 noun phrases by selecting the most relevant and important terms that co-occurred in at least eight articles to construct and visualize co-occurrence networks (Figure 3), in which terms are located such that the distance between two terms indicates their number of co-occurrences, and colors can be used to indicate the clusters of terms. Figure 3 indicates a clear distinction between basic and clinical research, with basic research located in the left part of the map (the blue and red terms, involving cells, molecules and animal related research areas) and clinical research in the right part (the green terms, related to human related research areas). Connections between basic research areas on the one hand and clinical research areas on the other hand are clearly visible in the term maps. Strong linkage exsits between basic and clinical terms, with particularly good examples being "cell" and "patient", "mouse" and "patient", indicating that clinicians can conduct basic research and biologists applied research and both can be interested in each other's research. The classification of such articles as mixed research is thus justified.

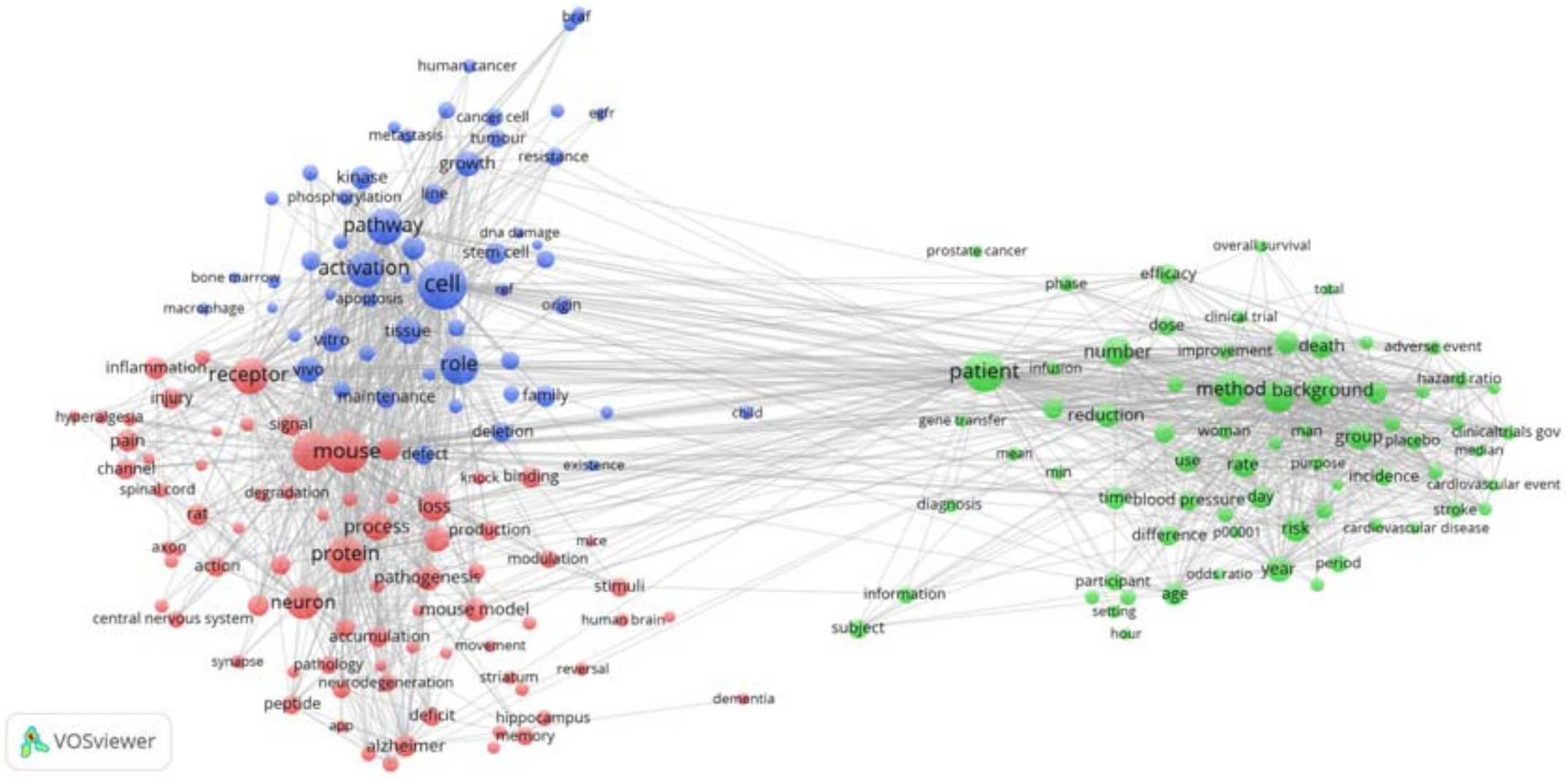

Figure 3. Co-occurrence map between "basic" terms (the blue and red terms) and "clinical" terms (the green terms)

Based on the above analysis of article labels, journals and terms, the classification into basic, clinical or mixed research from the perspective of recommenders' research areas and subjective concerns is reasonable, and is consistent with the objective and natural content patterns of articles.

*Association between citations and FFa scores for different research levels*

There was a similar low but significant correlation between FFa scores and citations for Basic Research (Spearman $r = 0.356$, $n = 18493$, $p < 0.01$), Clinical Research (Spearman $r = 0.287$, $n = 9069$, $p< 0.01$) and Mixed Research (Spearman $r = 0.402$, $n = 692$, $p < 0.01$).

Since both research level and the percentile interval of citations/FFa scores are categorical variables, the column proportions tests (*z*-test) rather than column means tests (*t*-test) were used to determine the distribution of publications assigned to each research level across six percentile intervals of citations/FFa scores (i.e. the "column proportions"). Then these proportions are compared among different research levels (based on the relative ranking) (see IBM Knowledge Center, SPSS Statistics 20.0.0, http://www-01.ibm.com/support/knowledgecenter/ SSLVMB_20.0.0/com .ibm.spss.statistics.help/sig_tests_colprop_ex.htm?lang=en). As is shown in Table 5, a subscript letter is assigned to the categories of the column variable, i.e., "A" for basic research, "B" for clinical research, and "C" for mixed research. For each pair of columns, the column proportions (for each row) were compared using a *z*-test. If a pair of values was significantly different, the values were assigned different subscript letters. For each significant pair, the Key Letter of the smaller category was placed under that of the larger category. For example, for the set of tests associated with the top 1%-0.1% of FFa scores, B is under A, and both A and B are under C, indicating that, for this FFa score class, the proportion of publications assigned to mixed research (6.4%) was greater than that assigned to basic research (0.9%), which in turn was greater than that assigned to clinical research (0.3%). Within the same class of citations, both the letter A and B are under C, but A and B are not reported as a significant pair, i.e., one is not placed under the other, implying that for papers with citations ranked in the top 1%-0.1% in our sample, mixed research (5.2%) represented a larger proportion than either basic research or clinical research, but basic research and clinical research did not differ significantly (each having a proportion of 0.8%).

Table 5. Column proportions test (three research levels vs. FFa Scores/Citations)

| | | Basic Research (A) | Clinical Research (B) | Mixed Research (C) |
|---|---|---|---|---|
| FFa Scores | above top 0.1% | 0.001 | 0.000 | 0.009 (A B) |
| | top 1%-0.1% | 0.009 (B) | 0.003 | 0.064 (A B) |
| | top 10%-1% | 0.076 (B) | 0.027 | 0.357 (A B) |
| | top 20%-10% | 0.148 (B) | 0.090 | 0.344 (A B) |
| | top 50%-20% | 0.281 (C) | 0.286 (C) | 0.150 |
| | below top 50% | 0.486 (C) | 0.594 (A C) | 0.077 |
| | total | 1 | 1 | 1 |
| Citations | above top 0.1% | 0.001 | 0.001 | 0.004 (A) |
| | top 1%-0.1% | 0.008 | 0.008 | 0.052 (A B) |
| | top 10%-1% | 0.096 (B) | 0.065 | 0.250 (A B) |
| | top 20%-10% | 0.112 (B) | 0.071 | 0.182 (A B) |
| | top 50%-20% | 0.327 (B) | 0.250 | 0.311 (B) |
| | below top 50% | 0.456 (C) | 0.604 (A C) | 0.201 |
| | total | 1 | 1 | 1 |

Significant at the 0.05 level (2-tailed). For each significant pair, the key of the smaller category is placed under the category with the larger proportion

The definition of mixed research forced this type of study typically to have at least 2 F1000 reviewers, so it has an advantage over the other two types because of the way that FFa scores are partly related to the number of reviewers. Those papers recommended by a single reviewer whose specialties span both basic and clinical areas, for instance, cardiovascular biology and cardiovascular disorders, are also assigned to mixed research. The percentage of articles with FFa Score of 1 (53/692, 7.7%) is much smaller for the mixed research level than the basic (8,983/18,493, 49%) or clinical (5,385/9,069, 59.4%) research levels. A mixed publication in general has at least two recommendations, which implies that mixed publications can be expected to have higher FFa scores than basic and clinical publications. Thus a comparison of the mixed research with the other two levels for FFa scores is biased. In addition, there is significant correlation between FFa scores and citations, a bias in FFa scores implies a bias in citations, i.e., publications with multiple recommendations and publications with higher FFa scores generally can also be expected to have more citations. To control the potential bias for FFa scores caused by the mixed research, only publications assigned to the basic or clinical research levels were included for further analysis (Table 6).

Table 6. Column proportions test (two research levels vs. FFa Scores/Citations)

| | | Basic Research (A) | Clinical research (B) |
|---|---|---|---|
| FFa Scores | above top 0.1% | 0.001 | 0.000 |
| | top 1%-0.1% | 0.009 (B) | 0.003 |
| | top 10%-1% | 0.076 (B) | 0.027 |
| | top 20%-10% | 0.148 (B) | 0.090 |
| | top 50%-20% | 0.281 | 0.286 |
| | below top 50% | 0.486 | 0.594 (A) |
| | total | 1 | 1 |
| Citations | above top 0.1% | 0.001 | 0.001 |
| | top 1%-0.1% | 0.008 | 0.008 |
| | top 10%-1% | 0.096 (B) | 0.065 |
| | top 20%-10% | 0.112 (B) | 0.071 |
| | top 50%-20% | 0.327 (B) | 0.250 |
| | below top 50% | 0.456 | 0.604 (A) |
| | total | 1 | 1 |

Significant at the 0.05 level (2-tailed). For each significant pair, the key of the smaller category is placed under the category with the larger proportion

In Table 5, we defined FFa scores or citations ranked in the top 20% as “excellent” and those in the bottom 50% as “common.” More of the “excellent” publications, whether measured by FFa scores or citations, were mixed research, with fewer being basic research and clinical research the fewest of all. The opposite was true for “common” publications, whether measured by FFa scores or citations: most were clinical research, then basic research, then mixed research. Excluding the mixed research level (Table 6), the proportion of publications assigned to the basic research level exceeds that assigned to the clinical research level for “excellent” articles, whether measured by FFa scores or citations. Thus no significant difference was found between FFa scores and citations in assessing publication excellence across research levels. We can conclude that research level has little impact on the internal consistency of the two measures.

An article can be assigned to only one research level, but can be tagged with more than one label. In the next section, we examine the effect of article type on the difference between FFa scores and citations.

*Association between citations and FFa scores for different article types*

F1000 reviewers assign labels to each article from a standard list when evaluating and rating biomedical articles. As shown in Table 7, the large majority (nearly 75%) of labeled articles were marked "New Finding," while 26% and 23% were categorized as "Interesting Hypothesis" and "Confirmation," respectively. "Technical Advance," "Controversial" and "Novel Drug Target" were assigned to 13%, 8% and 5% of articles, respectively. Few articles were labeled "Changes Clinical Practice" (1.8%) or "Refutation" (1.3%), with even fewer labeled "Good for Teaching" or "General Interest."

Within the assigned labels, little internal consistency exists between FFa scores and citations (Table 7). For example, the proportion of publications labeled "Clinical Trial: RCT" with top 10% citation rankings (3.5%) is significantly larger than for other article types (0.8%–1.3%), with the only exceptions being "Changes Clinical Practice" and "Reputation". However, a very different picture exists for FFa scores ranked in the top 10%. A large difference also exists between the top 1% of FFa scores and the top 1% of citations for publications with the "Interesting Hypothesis", "Technical Advance" and "Novel Drug Target" labels, the proportions of which exceed those labeled "New Finding", "Confirmation", "Review/Commentary" and "Non-RCT Clinical Trial" when measured by FFa score, yet are similar when measured by citations. The same difference exists in publications labeled "General Interest", "New Finding", "Confirmation", "Technical Advance", and so on. Article type seems to substantially impact the difference between FFa scores and citations.

According to the percentile distribution pattern, we divided the sample into four groups:_*Group 1*: high FFa scores and high citations (both in top 20%, *N*=2,506); *Group 2:* low FFa scores and low citations (both in bottom 50%, *N*=8,971); *Group 3*: high FFa scores and low citations (in top 20% and bottom 50%, respectively, N=1,332); and *Group 4*: low FFa and high citations (bottom 50% and in top 20%, respectively, N=1,621).

We then focused on the sub-sample of Groups 3 and 4, which represented inconsistent assessment results between FFa scores and citations. Only a few such article types are assigned as General Interest and Good for Teaching, which makes no sense for the conclusion. Using column proportions tests (Table 7), we found that *the non-primary research* (such as Review/Commentary) or *evidence-based research articles* (such as Systematic Review/Meta-Analysis, Clinical Trial: RCT, Clinical Trial: non-RCT, Confirmation, New Finding and Technical Advance) were more likely than other types to be highly cited yet less highly recommended (40%-70% vs. 10%-30%). By contrast, the *translational or transformative research articles* (e.g., Interesting Hypothesis, Controversial Topic, Novel Drug Target, Changes Clinical Practice and Refutation) were more likely to be highly rated by peer reviewers yet less cited than articles with other labels (60%-80% vs. 20%-50%).

Additionally, we have used a citation window of 3-5 years for articles published in 2010. In fact, for all articles included in this study published between 1999 and 2010, it is a citation window of 3-15 years. This means that articles from 1999 will have 15 years of citations and articles from 2010 will have 3-4 years of citations. Other factors being equal, articles from 1999 could be expected to have perhaps three or four times more citations than articles from 2010 in our dataset, which may bias the conclusions. In order to figure out whether the differences related to citations are due to changes of different article types over time, we reported the proportions of the 28254 articles classified as non-primary research or evidence-based research, and translational research or transformative research by year (Figure 4). Figure 4 showed that non-primary research or evidence-based research articles seem to be more common than translational research or transformative research articles for earlier years (80% versus 20%). Nevertheless, the percentage of the former research type showed a “slow decline—subtle rise” trend and the latter just the opposite. In general, the proportions of two research types have not witnessed a substantial change annually. They appear to be maintained at a relatively stable level, i.e., 75% and 25%, respectively. We could conclude that the use of a variable citation window (3-15 years) does not make a great difference on the above mentioned results.

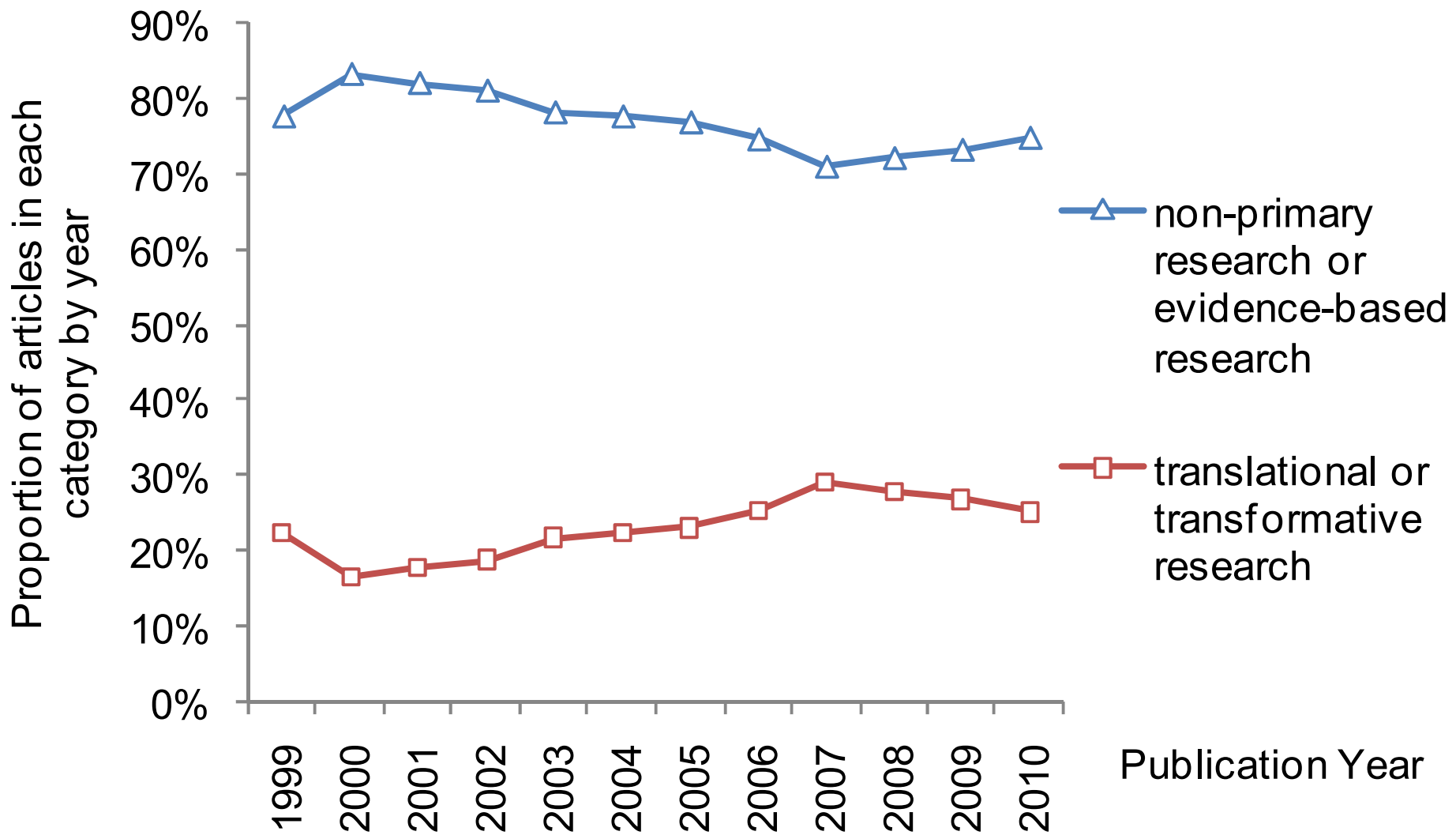

Figure 4. The proportions of articles in each category by year

Table 7. Column proportions test (article types v.s. Citations/FFa Scores)

| | Interesting Hypothesis (A) | New Finding (B) | Confirmation (C) | Technical Advance (D) | Controversial (E) | Novel Drug Target (F) | Clinical Trial: RCT (G) | Good for Teaching (H) | Review/ Commentary (I) | Clinical Trial: Non-RCT (J) | Changes Clinical Practice (K) | Systematic Review/ Meta-analysis (L) | Refutation (M) | General Interest (N) |
|---|---|---|---|---|---|---|---|---|---|---|---|---|---|---|
| ***N*** (total) | 7296 | 20937 | 6574 | 3807 | 2264 | 1505 | 1394 | 52 | 830 | 1068 | 495 | 44 | 354 | 11 |
| **Citations** | | | | | | | | | | | | | | |
| above Top 0.1% | 0.001 | 0.001 | 0.002 | 0.003 | 0.000 | 0 (.a) | 0.004 (B) | 0 (.a) | 0.002 | 0.002 | 0 (.a) | 0 (.a) | 0.003 | 0 (.a) |
| Top 1%~0.1% | 0.012 | 0.010 | 0.013 | 0.017 (B) | 0.012 | 0.009 | 0.035 (A B C D E F I J) | 0.019 | 0.011 | 0.008 | 0.051 (A B C D E F I J) | 0 (.a) | 0.020 | 0.091 |
| Top 10%~1% | 0.112 (B) | 0.099 | 0.102 | 0.125 (B C J) | 0.115 | 0.106 | 0.166 (A B C D E F I J) | 0.096 | 0.102 | 0.080 | 0.17 (A B C F I J) | 0.159 | 0.099 | 0.091 |
| Top 20%~10% | 0.113 | 0.111 | 0.102 | 0.122 (J) | 0.109 | 0.109 | 0.131 (C J) | 0.115 | 0.099 | 0.081 | 0.087 | 0.136 | 0.071 | 0.091 |
| Top 50%~20% | 0.308 | 0.32 (C) | 0.299 | 0.301 | 0.294 | 0.304 | 0.280 | 0.231 | 0.281 | 0.318 | 0.255 | 0.364 | 0.302 | 0.182 |
| below Top 50% | 0.455 (G) | 0.459 (G) | 0.483 (A B D G) | 0.432 | 0.47 (G) | 0.471 (G) | 0.383 | 0.538 | 0.505 (D G) | 0.511 (A B D G) | 0.438 | 0.341 | 0.506 (G) | 0.545 |
| **FFa Scores** | | | | | | | | | | | | | | |
| above Top 0.1% | 0.003 (B) | 0.001 | 0.002 | 0.004 (B) | 0.003 | 0.005 (B) | 0.001 | 0 (.a) | 0 (.a) | 0 (.a) | 0 (.a) | 0 (.a) | 0.008 (B) | 0.091 (A B C D E F G) |
| Top 1%~0.1% | 0.022 (B C I J) | 0.010 | 0.014 | 0.022 (B C I J) | 0.019 (B I J) | 0.026 (B C I J) | 0.013 | 0 (.a) | 0.001 | 0.003 | 0.03 (B I J) | 0 (.a) | 0.023 (I J) | 0.091 (I J) |
| Top 10%~1% | 0.132 (B C G I J) | 0.085 (I J) | 0.086 (I J) | 0.137 (B C G I J) | 0.13 (B C G I J) | 0.139 (B C G I J) | 0.077 (I J) | 0.096 | 0.022 | 0.037 | 0.166 (B C G I J) | 0.114 (I) | 0.107 (I J) | 0.364 (G I J) |
| Top 20%~10% | 0.185 (B C I J) | 0.158 (C I J) | 0.138 (I J) | 0.179 (B C I J) | 0.189 (B C I J) | 0.191 (B C I J) | 0.165 (I J) | 0.25 (I J) | 0.073 | 0.086 | 0.21 (C I J) | 0.068 | 0.184 (I J) | 0.182 |
| Top 50%~20% | 0.267 | 0.284 (A C) | 0.260 | 0.261 | 0.270 | 0.279 | 0.307 (C) | 0.423 | 0.339 (A B C D E J) | 0.265 | 0.362 (A B C D E F J) | 0.182 | 0.302 | 0.273 |
| below Top 50% | 0.391 (K) | 0.461 (A D E F K) | 0.501 (A B D E F G H K M) | 0.397 (K) | 0.389 (K) | 0.36 (K) | 0.436 (F K) | 0.231 | 0.565 (A B C D E F G H K M) | 0.61 (A B C D E F G H K M) | 0.232 | 0.636 (F H K) | 0.376 (K) | 0 (.a) |
| **Two opposite groups of Citations and FFa Scores** | | | | | | | | | | | | | | |
| *N* (only Group 3 and 4) | 823 | 2246 | 647 | 451 | 259 | 146 | 177 | 8 | 107 | 108 | 51 | 13 | 41 | 3 |
| low FFa Scores but high citations | 0.352 | **0.502 (A E F M)** | **0.556 (A D E F K M)** | **0.448 (A E F M)** | 0.301 | 0.267 | **0.701 (A B C D E F K M)** | 0.25 | **0.766 (A B C D E F K M)** | **0.722 (A B C D E F K M)** | 0.314 | **0.769 (E F M)** | 0.146 | 0 (.a) |
| high FFa Scores but low citations | **0.648 (B C D G I J L)** | 0.498 (G I J) | 0.444 (G I J) | 0.552 (C G I J) | **0.699 (B C D G I J L)** | **0.733 (B C D G I J L)** | 0.299 | 0.75 | 0.234 | 0.278 | **0.686 (C G I J L)** | 0.231 | **0.854 (B C G I J L)** | 1 (.a) |

Significant at the 0.05 level (2-tailed). For each significant pair, the key of the smaller category is placed under the category with the larger proportion. a. The column proportion is 0 or 1, not for compare; b. Bonferroni adjustments are used to adjust the significance values.

## Discussion and Conclusions

Few empirical studies have looked at the main reasons for the differences between citation analysis and peer review. In this paper, we investigated the effect of research level and article type on the consistency of assessments by citations and FFa scores. We found that research level has little impact on the internal consistency of these two measures. In general, whether measured by citations or FFa scores, significantly more articles assigned to mixed research scored highly than those assigned to basic research, which scored higher than clinical research articles. This result, to some extent, is in agreement with the finding of van Eck et al., (2013)'s finding that low-impact research areas tend to focus on clinical intervention research, while high-impact research areas are often more oriented to basic and diagnostic research. It seems that, when compared with the absolute number of such measures as citations and FFa scores, both citation analysis and peer review may underestimate the impact of clinical research as compared to basic research. When taking article types into account, only one study examined the effect of F1000 labels on the association between FFa scores and citations (Mohammadi & Thelwall, 2013). That study took a random sample of F1000 medical articles and found that citation counts and FFa scores were significantly different for only two types of articles, namely "New Finding" and "Changes Clinical Practice". FFa scores value the appropriateness of medical research for clinical practice and "New Finding" articles are more highly cited. These conclusions were based on a small-scale sample which was limited to medical articles only. In our study, we analyzed a larger-scale and more strictly predefined sample of publications from three pairs of research fields representing the biological and corresponding medical research within the same disease area. We found that article type, as assigned by F1000 faculty members, has a substantial effect on the difference between FFa scores and citations. The two measures represent significantly different things for the two groups of article types. Articles in the first group are more highly cited rather than highly recommended. These mainly include non-primary research, such as review/commentary and systematic review/meta-analysis, and evidence based research, such as RCTs, confirmation, new finding and technical advances. The use of a variable citation window (3-15 years) does not greatly affect this result. Publications in the second group, however, are more highly recommended but gather relatively fewer citations. This group mainly involves either translational research (such as research indicating novel drug targets or with the potential for changing clinical practice) or transformative research (such as interesting hypotheses, controversial topics and refutation of currently held beliefs). FFa scores appear to be more conducive than citations to identifying important articles for clinical and translational research, transformative research and high risk research. This difference can be expected because academic authors are the ones who cite previous studies while an article's potential for scientific revolution or changing clinical practice is evaluated from the perspective of practitioners.

But both citations and FFa scores in research evaluation have innate limitations in research evaluation. In a recent editorial, Marks, Marsh, Schroer, and Stevens (2013) noted an alarming trend within the biological/biomedical research literature toward the citation of review articles rather than the primary research papers in which the original findings were described. Others have also pointed out that reviews, or secondary literature, are more cited than the original research papers. Patsopoulos, Analatos, and Ioannidis (2005) measured the citation impact of articles using various study designs—including meta-analyses, RCTs, cohort studies, case-control studies, case reports, nonsystematic reviews and decision analysis or cost-effectiveness analysis—and found that the citation impact of various study designs follows the order proposed by most current theoretical hierarchies of evidence. Overall, meta-analyses received more citations than any other study design both in the first two years after publication and in the longer term, RCTs become the second most cited study design and case reports have negligible impact. Knottnerus and Knottnerus (2009) also found that systematic reviews currently attract many more citations than the original studies. Our study is commensurate with and further validates the above mentioned arguments: the secondary literature and the research evidence are more cited. No wonder several excellent journals such as *Molecular Biology of the Cell*, *EMBO Journal* and *Traffic* have taken action to recognize the contribution of original research more appropriately, e.g., instructing authors to cite the primary literature and to limit the citation of review articles, and asking referees to note when too many reviews are cited in preference to original publications. The San Francisco DORA also recommended that researchers "wherever appropriate, cite primary literature in which observations are first reported rather than reviews in order to give credit where credit is due." In addition, since citation is one type of measure which increases with time, and different research fields have different citation speeds and citation durability; in addition, individual papers vary in their aging patterns (Costas, Van Leeuwen, & Van Raan, 2010; Van Raan, 2004; Wang, 2013). These elements must be taken into account when using citation analysis for research evaluations and funding allocations, and thus micro-level analysis (and especially analysis of the individual level) is one of the most difficult and problematic levels of analysis in bibliometric studies.

FFa scores in the F1000 system also have obvious limitations. First, while making citations is relatively easy, recommendations are more time-consuming and therefore less numerous, leading to many ties and thus reducing their usefulness for ranking articles. Second, the large numbers of citations clearly produces a reliable source of data; however, the number of recommendations is much smaller, making it a less reliable data source than citations. The limited availability of recommendations has to do with their reliability. Additionally, the academic position of F1000 experts is unknown despite the importance of the academic status of reviewers, and the points scheme used by F1000 to convert judgments into a score is relatively arbitrary, such that the article ranking varies according to the details of the scheme. Finally, subjective assessments of the merit and likely impact of scientific publications often manifest biases in the form of low inter-rater reliability. Using two large datasets, including F1000 and a Wellcome Trust grant panel, in which scientists make qualitative assessments of scientific merit, Eyre-Walker and Stoletzki (2013) showed that scientists' judgments are strongly influenced by the journal in which a paper is published and cannot consistently and independently judge the merit of a paper, or predict its future impact as measured by citations. Bornmann (2014) also argued that analyses of the reliability of the F1000Prime peer review system showed low agreement among faculty members. Despite its problems, the subjective assessment of research by experts has always been considered a gold standard, and has been championed by researchers and funders alike. Yet the analysis mentioned above now raises serious questions about this process (Eisen, MacCallum, & Neylon, 2013).

Although imperfect, citation and recommendation are still important tools for scientific recognition. The data and indicators provided by F1000 are undoubtedly rich and valuable, and this tool has strong potential for research evaluation, being a good complement to alternative metrics for research assessments. Furthermore, since most studied articles were reviewed in the year of publication (Waltman & Costas, 2014), F1000 can be useful when quick evaluations are needed. However, compared with citations, F1000 fails to consistently and dynamically identify the most important publications in a long time window.

One limitation of this study is that we used the specialty information of faculty members in F1000 to assign articles to the basic, clinical or mixed research levels. F1000 reviewers and the number of reviewed papers were not distributed uniformly across the disciplines. A paper can be reviewed by faculty members specialized in several subjects, including those with specialties not covered by the six subjects included in this study, which may cause biases in the classification of publications according to research levels. Currently no consensus exists regarding the methodology used to classify publications into research levels along the basic-to-applied spectrum of research (Boyack, Patek, Ungar, Yoon, & Klavans, 2014), especially with the increase in university-government-industry partnerships in biomedical sciences (Campbell, Powers, Blumenthal, & Biles, 2004). Nevertheless, we are confident that, as a quantitative study on the effect of research level and article type on the differences between citation analysis and peer reviews, our study has achieved its purpose: to answer questions about the degree of correlation of recommendations with citations, and whether recommendations capture something different from citations.

Attempts to improve the application of bibliometric tools in research assessment procedures should consider the proportions of three types of publications: (1) evidence based research publications; (2) transformative research or high risk research publications; and (3) translational research publications. Further analysis that considers these three types of publications might reveal new and useful information, which may help more accurately determine the relative value of different scholarly contributions.

**Acknowledgment**

This study was supported by the National Natural Science Foundation of China (Grant No. 71373252) and the Institute of Medical Information of the Chinese Academy of Medical Sciences (Grant No. 14R0106)